\documentclass[pre,aps,amssymb,amsmath,superscriptaddress,twocolumn]{revtex4}
\usepackage{amsmath}
\usepackage{amssymb}
\usepackage{epsfig}
\usepackage{amscd}
\usepackage{graphicx,psfrag,xspace}
\usepackage{float}
\usepackage[normalem]{ulem}

\begin{document}

\title{Counting statistics: a Feynman-Kac perspective}
\author{A. Zoia}
\email{andrea.zoia@cea.fr}
\affiliation{CEA/Saclay, DEN/DANS/DM2S/SERMA/LTSD, 91191 Gif-sur-Yvette, France}
%\affiliation{Commissariat \`a l'Energie Atomique et aux Energies Alternatives, Direction de l'Energie Nucl\'eaire, D\'epartement de Mod\'elisation des Syst\`emes et Structures, Service d'Etudes des R\'eacteurs et de Math\'ematiques Appliqu\'ees, CEA/Saclay, 91191 Gif-sur-Yvette, France}
\author{E. Dumonteil}
\affiliation{CEA/Saclay, DEN/DANS/DM2S/SERMA/LTSD, 91191 Gif-sur-Yvette, France}
\author{A. Mazzolo}
\affiliation{CEA/Saclay, DEN/DANS/DM2S/SERMA/LTSD, 91191 Gif-sur-Yvette, France}

\begin{abstract}
By building upon a Feynman-Kac formalism, we assess the distribution of the number of hits in a given region for a broad class of discrete-time random walks with scattering and absorption. We derive the evolution equation for the generating function of the number of hits, and complete our analysis by examining the moments of the distribution, and their relation to the walker equilibrium density. Some significant applications are discussed in detail: in particular, we revisit the gambler's ruin problem and generalize to random walks with absorption the arcsine law for the number of hits on the half-line.
\end{abstract}
\maketitle

\section{Introduction}
\label{introduction}

Many physical processes can be described in terms of a random walker evolving in the phase space~\cite{hughes, weiss, bouchaud_desorder, avraham}, and one is often interested in assessing the portion of time $t_V$ that the system spends in a given region $V$ of the explored space, when observed up to time $t$~\cite{redner, condamin_benichou, condamin, benichou_grebenkov, barkai, klafter, grebenkov, grebenkov_jsp, majumdar_occupation}. This is key to understanding the dynamics of radiation transport, gas flows, research strategies or chemical/biological species migration in living bodies, only to mention a few, the time spent in $V$ being proportional to the interaction of particle with the target medium~\cite{cercignani, wigner, jacoboni_book, gamma_biology, lecaer, schlesinger, zoia}.

For Brownian motion, a celebrated approach to characterizing the probability density of the residence time $t_V$ has been provided by Kac (based on Feynman path integrals) in a series of seminal papers, and later extended to Markov continuous-time processes~\cite{kac_original, kac_berkeley, kac_darling, kac}. For a review, see for instance~\cite{majumdar_review}. Feynman-Kac formalism basically allows writing down the evolution equation for the moment generating function of $t_V$ for arbitrary domains, initial conditions and displacement kernels. This approach has recently attracted a renovated interest~\cite{agmon_original, berezhkovskii, benichou_epl, agmon_lett, agmon, zoia_dumonteil_mazzolo, zdm_prl, zdm_pre}, and has been also extended to non-Markovian processes~\cite{barkai, turgeman, barkai_jsp}. As a particular case, imposing leakage boundary conditions leads to the formulation of first-passage problems~\cite{redner, condamin_benichou, klafter_fptd}. However, for those physical systems that are intrinsically discrete, the natural variable is the number of hits $n_V$ in $V$ when the process is observed up to the $n$-th step, rather than time $t_V$~\cite{delmoral, zdm_pre, pitman, iosifescu, csaki}. When $n_V$ is large, we can approximate the number of events falling in $V$ by $n_V \propto t_V$ (the so-called diffusion limit), but this simple proportionality breaks down when $V$ is small with respect to the typical step size of the walker, and/or the effects of absorption are not negligible, so that the diffusion limit is not attained~\cite{blanco, zdm_prl}.

In this paper, we derive a discrete Feynman-Kac equation for the evolution of the probability generating function of $n_V$ for a broad class of stochastic processes with scattering and absorption, and we illustrate this approach by explicitly working out calculations for some significant examples, such as the gambler's ruin problem, or the arcsine law. For the arcsine law, in particular, the Feynman-Kac formulae allow generalizing the well-known Sparre Andersen results to random walks with absorption. Our analysis of the counting statistics is then completed by examining the moments of $n_V$, which can also be obtained by building upon the Feynman-Kac formalism, and their asymptotic behavior when $n$ is large. In particular, we show that the asymptotic moments can be expressed as a function of the particle equilibrium distribution, which generalizes analogous results previously derived in terms of survival probabilities~\cite{zdm_prl}.

This paper is organized as follows. In Sec.~\ref{feynman_kac_eq} we introduce a discrete Feynman-Kac formula for a class of random walkers with scattering and absorption. Then, in Sec.~\ref{applications_gf} we discuss some applications where the generating function can be explicity inverted to give the probability of the number of hits. In Sec.~\ref{moments_formulae} we extend our analysis to the moments of $n_V$, and in Sec.~\ref{applications_moments} we examine some examples of moment formulae. A short digression on the diffusion limit is given in Sec.~\ref{diffusion}. Perspectives are finally discussed in Sec.~\ref{conclusions}.

\section{Feynman-Kac equations}
\label{feynman_kac_eq}

Consider the random walk of a particle starting from an isotropic point-source ${\cal S}({\mathbf r}|{\mathbf r}_0)=\delta({\mathbf r}-{\mathbf r}_0)$ located at ${\mathbf r}_0$. At each collision, the particle can be either scattered with probability $p_s$, or absorbed with probability $p_a=1-p_s$ (in which case the trajectory terminates). We introduce the quantity $T({\mathbf r'}\to{\mathbf r})$, namely, the probability density of performing a displacement from ${\mathbf r'}$ to ${\mathbf r}$, between any two collisions~\cite{spanier, lux}. For the sake of simplicity, we assume that scattering is isotropic, and that displacements are equally distributed.

Suppose that a particle emitted from ${\mathbf r}_0$ is observed up to entering the $n$-th collision. Our aim is to characterize the distribution $P_n(n_V|{\mathbf r}_0)$, where $n_V$ is the number of collisions in a domain $V$. We can formally define
\begin{equation}
n_V(n)=\sum^{n}_{k=1} V({\mathbf r}_k),
\label{coll_num_definition}
\end{equation}
where $V({\mathbf r}_k)$ is the marker function of the region $V$, which takes the value $1$ when the point ${\mathbf r}_k \in V$, and vanishes elsewhere. We adopt here the convention that the source is not counted, i.e., the sum begins at $k=1$. Clearly, $n_V$ is a stochastic variable, depending on the realizations of the underlying process, and on the initial condition ${\mathbf r}_0$. The behavior of its distribution, $P_n(n_V|{\mathbf r}_0)$, is most easily described in terms of the associated probability generating function
\begin{equation}
F_n(u|{\mathbf r}_0)= \langle u^{n_V}\rangle_n({\mathbf r}_0) = \sum^{+\infty}_{n_V=0}P_n(n_V|{\mathbf r}_0)u^{n_V},
\end{equation}
which can be interpreted as the discrete Laplace transform (the transformed variable being $u$) of the collision number distribution. The derivation of an evolution equation for $F_n(u|{\mathbf r}_0)$ is made simpler if we initially consider trajectories starting with a particle entering its first collision at ${\mathbf r}_1$. Random flights are semi-Markovian (i.e., Markovian at collision points), which allows splitting the trajectory into a first jump, from ${\mathbf r}_1 $ to ${\mathbf r}_1 + \Delta$ (the displacement $\Delta$ obeying the jump length density $T$), and then a path from ${\mathbf r}_1 + \Delta$ to ${\mathbf r}_n$, conditioned to the fact that the collision at ${\mathbf r}_1$ was a scattering event. If the collision is an absorption, the trajectory ends and there will be no further events contributing to $n_V$. This leads to the following equation for the generating function
\begin{eqnarray}
\tilde{F}_{n+1}(u|{\mathbf r}_1)
=u^{V({\mathbf r}_1)}\left[ p_s\langle \tilde{F}_n(u|{\mathbf r}_1 + \Delta)\rangle +p_a \right] ,
\label{equation_F_one}
\end{eqnarray}
where expectation is taken with respect to the random displacement $\Delta$. The tilde is used to recall that we are considering trajectories starting with a single particle entering the first collision at ${\mathbf r}_1$. We make then use of the discrete Dynkin's formula
\begin{equation}
\langle f({\mathbf r}_1+ \Delta)\rangle = \int T^*({\mathbf r}'\to {\mathbf r}_1) f({\mathbf r}')d{\mathbf r}',
\end{equation}
where $f$ is any sufficiently well behaved function and $T^*({\mathbf r}'\to {\mathbf r})$ is the adjoint kernel associated to  $T({\mathbf r}'\to {\mathbf r})$~\cite{meyn}. We therefore obtain the discrete Feynman-Kac equation
\begin{eqnarray}
\tilde{F}_{n+1}(u|{\mathbf r}_1)
=u^{V({\mathbf r}_1)}\left[ p_s\int  T^*({\mathbf r}'\to {\mathbf r}_1)\tilde{F}_n(u|{\mathbf r}')d{\mathbf r}' +p_a \right],
\label{equation_F}
\end{eqnarray}
with the initial condition $\tilde{F}_1(u|{\mathbf r}_1)=u^{V({\mathbf r}_1)}$. Finally, by observing that the first collision coordinates ${\mathbf r}_1$ obey the probability density $T({\mathbf r}_0\to{\mathbf r}_1)$, it follows
\begin{equation}
F_n(u|{\mathbf r}_0)=\int \tilde{F}_n(u|{\mathbf r}_1) T({\mathbf r}_0\to{\mathbf r}_1)d{\mathbf r}_1.
\label{feynman_x0}
\end{equation}
Knowledge of $F_{n}(u|{\mathbf r}_0)$ allows explicitly determining $P_n(n_V|{\mathbf r}_0)$. Indeed, by construction the probability generating function $F_{n}(u|{\mathbf r}_0)$ is a polynomial in the variable $u$, the coefficient of each power $u^k$ being $P_n(n_V=k|{\mathbf r}_0)$. In particular, it follows that the probability that the particles never touch (or come back to, if the source ${\mathbf r}_0 \in V$) the domain $V$ is obtained by evaluating $F_{n}(u|{\mathbf r}_0)$ at $u=0$, i.e., $P_n(0|{\mathbf r}_0) = F_{n}(0|{\mathbf r}_0)$.

\begin{figure}[t]
   \centerline{\epsfclipon \epsfxsize=9.0cm
\epsfbox{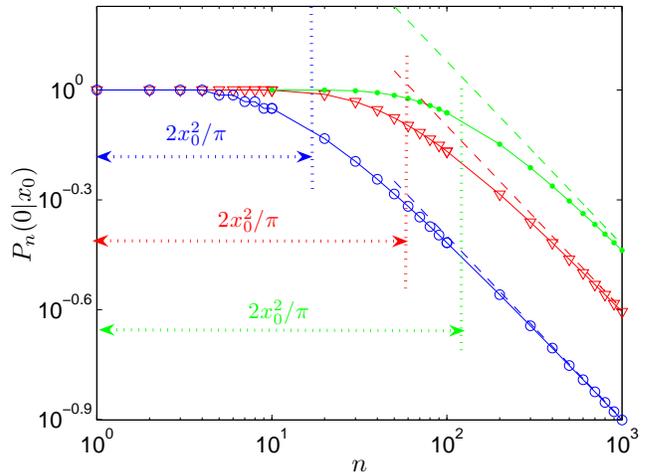} }
   \caption{(Color online). The probability $P_n(0|x_0)$ that the gambler is not ruined at the $n$-th bet, given an initial capital $x_0$. Bets are modeled by discrete random increments of fixed size $s = \pm 1$. Blue circles: $x_0=5$; red triangles: $x_0=10$; green dots: $x_0=15$. Lines have been added to guide the eye. Dashed lines correspond to the asymptotic result $P_n(0|x_0) \simeq \sqrt{2/\pi} x_0 n^{-1/2}$. The interval $2x_0^2/\pi$ is also shown for each $x_0$.}
   \label{fig1}
\end{figure}

\section{Collision number distribution: examples of calculations}
\label{applications_gf}

Direct calculations based on the discrete Feynman-Kac formulae, Eqs.~\eqref{equation_F} and~\eqref{feynman_x0}, are in some cases amenable to exact results concerning $P_n(n_V|{\mathbf r}_0)$, at least for simple geometries and displacement kernels. In this Section, we shall discuss some relevant examples.

\subsection{The gambler's ruin}

\begin{figure}[t]
   \centerline{\epsfclipon \epsfxsize=9.0cm
\epsfbox{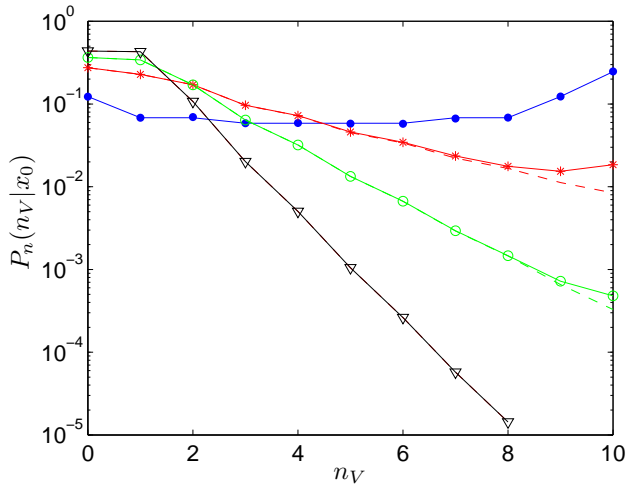} }
   \caption{(Color online). The arcsine law $P_n(n_V|x_0)$ with discrete jump lengths, as a function of $n_V$. The starting point is $x_0=0$, and $n=10$. Blue dots: $p_s=1$ (dots); red stars: $p_s=3/4$; green circles: $p_s=1/2$; black triangles: $p_s=1/4$. Lines have been added to guide the eye. Dashed curves are the asymptotic Eq.~\eqref{asy_disc} for the correponding value of $p_s$.}
   \label{fig2}
\end{figure}

Consider a gambler whose initial amount of money is $x_0 \ge 0$. At each (discrete) time step, the gambler either wins or loses a fair bet, and his capital increases or decreases, respectively, by some fixed quantity $s$ with equal probability. One might be interested in knowing the probability that the gambler is not ruined (i.e., that his capital has not reached zero, yet) at the $n$-th bet, starting from the initial capital $x_0$. This well-known problem~\cite{feller} can be easily recast in the Feynman-Kac formalism by setting a particle in motion on a straight line, starting from $x_0$, with scattering probability $p_s=1$ and a discrete displacement kernel $T(x' \to x)=\delta(x-x'-s)/2+\delta(x-x'+s)/2$. Setting $s=1$ amounts to expressing the capital $x_0$ in multiple units of the bet, and entails no loss of generality. The counting condition is imposed by assuming a Kronecker delta $V(x)=\delta_{x,0}$ in Eq.~\eqref{equation_F}: since the walker can not cross $x=0$ without touching it, solving the resulting equation for the quantity $F_{n}(0|x_0)$ gives therefore the required probability that the gambler is not ruined at the $n$-th bet. Now, we integrate Eq.~\eqref{equation_F} and use Eq.~\eqref{feynman_x0}: identifying the coefficient of the zeroth order term in the polynomial yields then
\begin{equation}
P_n(0|0)=\left\lbrace 1,\frac{1}{2},\frac{2}{4},\frac{3}{8},\frac{6}{16},\frac{10}{32},\frac{20}{64},\frac{35}{128},...\right\rbrace 
\end{equation}
for $n \ge 1$, when $x_0=0$. This is easily recognized as being the series
\begin{equation}
P_n(0|0)= \binom{n-1}{\lceil\frac{n-1}{2}\rceil}2^{1-n},
\end{equation}
$\lceil \cdot \rceil$ denoting the integer part. For $x_0=1,2,3,...$ we obtain for the first terms in the series
\begin{eqnarray}
P_n(0|1)=\left\lbrace \frac{1}{2},\frac{2}{4},\frac{3}{8},\frac{6}{16},\frac{10}{32},\frac{20}{64},\frac{35}{128},\frac{70}{256},...\right\rbrace  \nonumber \\
P_n(0|2)=\left\lbrace \frac{2}{2},\frac{3}{4},\frac{6}{8},\frac{10}{16},\frac{20}{32},\frac{35}{64},\frac{70}{128},\frac{126}{256},...\right\rbrace  \nonumber \\
P_n(0|3)=\left\lbrace \frac{2}{2},\frac{4}{4},\frac{7}{8},\frac{14}{16},\frac{25}{32},\frac{50}{64},\frac{91}{128},\frac{182}{256},...\right\rbrace  \nonumber \\
P_n(0|4)=\left\lbrace \frac{2}{2},\frac{4}{4},\frac{8}{8},\frac{15}{16},\frac{30}{32},\frac{56}{64},\frac{112}{128},\frac{210}{256},...\right\rbrace
\end{eqnarray}
After some rather lengthy algebra, by induction one can finally recognize the formula
\begin{equation}
P_n(0|x_0)= \sum_{k=0}^{\lceil(n+x_0-1)/2\rceil}\left[  \binom{n}{k} - \binom{n}{k-x_0} \right] 2^{-n}.
\label{gambler_eq}
\end{equation}
The quantity $P_n(0|x_0)$ is displayed in Fig.~\ref{fig1} as a function of $n$ for a few values of $x_0$. The larger the initial capital $x_0$, the longer $P_n(0|x_0) \simeq 1$ before decreasing. At large $n$, taking the limit of Eq.~\eqref{gambler_eq} leads to the scaling $P_n(0|x_0) \simeq \sqrt{2/\pi} x_0 n^{-1/2}$, in agreement with the findings in~\cite{majumdar_fptd}. This means that asymptotically the gambler is almost sure not to be ruined, yet, up to $n \simeq 2 x_0^2 / \pi $ bets. Note that Eq.~\eqref{gambler_eq} is the survival probability of the gambler: the first passage probability $W_n(0|x_0)$, i.e., the probability that the gambler is ruined exactly at the $n$-th bet, can be obtained from $W_n(x_0)=P_{n-1}(0|x_0)-P_{n}(0|x_0)$. As a particular case, for $0 < x_0 \le n$ and $n+x_0$ even, we recover the result in~\cite{feller}, namely
\begin{equation}
W_n(x_0)=\frac{x_0}{2^{n}n}\binom{n}{\frac{n+x_0}{2}}.
\end{equation}

\begin{figure}[t]
   \centerline{\epsfclipon \epsfxsize=9.0cm
\epsfbox{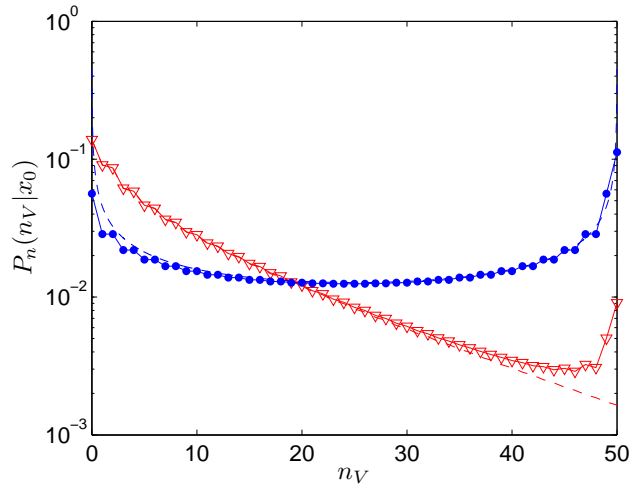} }
   \caption{(Color online). The arcsine law $P_n(n_V|x_0)$ with discrete jump lengths, as a function of $n_V$. The starting point is $x_0=0$, and $n=50$. Blue dots: $p_s=1$; dashed line: the asymptotic distribution $1/\sqrt{n_V(n-n_V)}\pi$. Red triangles: $p_s=0.95$; dashed asymptotic Eq.~\eqref{asy_disc}.}
   \label{fig3}
\end{figure}

\subsection{The arcsine law with discrete jumps}

Consider a walker on a straight line, starting from $x_0$. We are interested in assessing the distribution $P_n(n_V|x_0)$ of the number of collisions $n_V$ that the walker performs at the right of the starting point, when observed up to the $n$-th collision. This amounts to choosing $V(x)=H(x-x_0)$, $H$ being the Heaviside step function, in Eq.~\eqref{equation_F}. To fix the ideas, without loss of generality we set $x_0=0$, and we initially assume that $p_s=1$, i.e., the walker can not be absorbed along the trajectory. This is a well known and long studied problem, for both Markovian and non-Markovian processes~\cite{feller, majumdar_occupation, barkai, lamperti, majumdar_potential, godreche, baldassarri}: for Brownian motion, the average residence time in $V$ is simply $\langle t_V \rangle_t=t/2 $, whereas $t_V$ itself is known to obey the so called L\'evy's arcsine law $P_t(t_V) = 1/\sqrt{t_V(t-t_V)}\pi$, whose U shape basically implies that the particle will most often spend its time being always either on the positive or negative side of the axis~\cite{feller, watanabe, majumdar_fptd}. This counterintuitive result has been shown to asymptotically hold also for discrete-time random walks without absorption, for which one has $P_n(n_V|0) \simeq 1/\sqrt{n_V(n-n_V)}\pi$ when $n$ and $n_V$ are large (see for instance~\cite{majumdar_fptd}).

\begin{figure}[t]
   \centerline{\epsfclipon \epsfxsize=9.0cm
\epsfbox{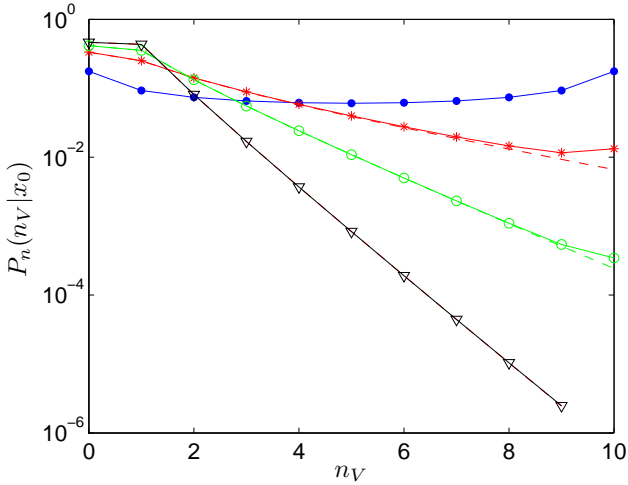} }
   \caption{(Color online). The arcsine law $P_n(n_V|x_0)$ with exponential jump lengths, as a function of $n_V$. The starting point is $x_0=0$, and $n=10$. Blue dots: $p_s=1$; red stars: $p_s=3/4$; green circles: $p_s=1/2$; black triangles: $p_s=1/4$. Lines have been added to guide the eye. Dashed curves are the asymptotic Eq.~\eqref{asy_cont} for the correponding value of $p_s$.}
   \label{fig4}
\end{figure}

The Feynman-Kac approach allows explicitly deriving $P_n(n_V|0)$. Again, assume a displacement kernel with discrete jumps $T(x' \to x)=\delta(x-x'-s)/2+\delta(x-x'+s)/2$, with $s=1$. Then, by integrating Eq.~\eqref{equation_F} and subsequently using Eq.~\eqref{feynman_x0} we compute the coefficients of the polynomial, which can be organized in an infinite triangle, whose first terms read\\

    \begin{tabular}{ c| c c c c c c c c}
     $n$ & $n_V=0$ & $1$ & $2$ & $3$ & $4$ & $5$ & $6$ & $7$\\
     \hline
     $0$ & $1$ & & & & & & &\\
     $1$ & $\frac{1}{2}$ & $\frac{1}{2}$ & & & & & &\\
     $2$ & $\frac{1}{4}$ & $\frac{1}{4}$ & $\frac{2}{4}$ & & & & &\\
     $3$ & $\frac{2}{8}$ & $\frac{1}{8}$ & $\frac{2}{8}$ & $\frac{3}{8}$ & & & &\\
     $4$ & $\frac{3}{16}$ & $\frac{2}{16}$ & $\frac{2}{16}$ & $\frac{3}{16}$ & $\frac{6}{16}$ & & &\\
     $5$ & $\frac{6}{32}$ & $\frac{3}{32}$ & $\frac{4}{32}$ & $\frac{3}{32}$ & $\frac{6}{32}$ & $\frac{10}{32}$ & &\\
     $6$ & $\frac{10}{64}$ & $\frac{6}{64}$ & $\frac{6}{64}$ & $\frac{6}{64}$ & $\frac{6}{64}$ & $\frac{10}{64}$ & $\frac{20}{64}$ &\\
     $7$ & $\frac{20}{128}$ & $\frac{10}{128}$ & $\frac{12}{128}$ & $\frac{9}{128}$ & $\frac{12}{128}$ & $\frac{10}{128}$ & $\frac{20}{128}$ & $\frac{35}{128}$\\
    \end{tabular}
\vspace{.5cm}

\begin{figure}[t]
   \centerline{\epsfclipon \epsfxsize=9.0cm
\epsfbox{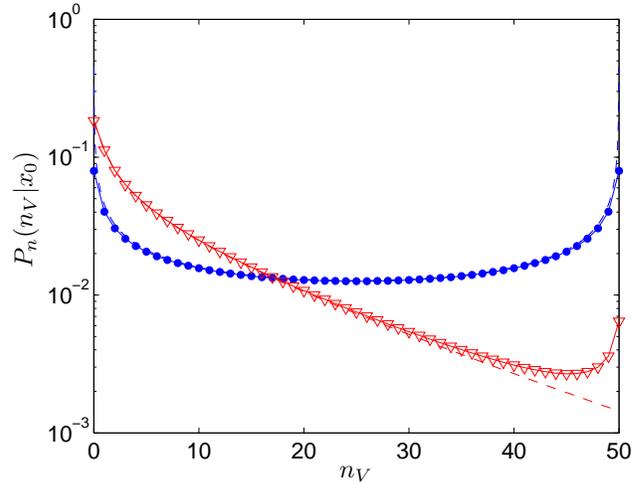} }
   \caption{(Color online). The arcsine law $P_n(n_V|x_0)$ with exponential jump lengths, as a function of $n_V$. The starting point is $x_0=0$, and $n=50$. Blue dots: $p_s=1$; dashed line: the asymptotic distribution $1/\sqrt{n_V(n-n_V)}\pi$. Red triangles: $p_s=0.95$; dashed line: asymptotic Eq.~\eqref{asy_cont}.}
   \label{fig5}
\end{figure}

Observe that this result is independent of $s$. To identify the elements $P_n(n_V|0)$, we initially inspect the column $n_V=1$ of the triangle, and recognize the underlying series as being given by terms of the kind $\binom{k}{\lceil k/2 \rceil}2^{-k}$. Then we realize that columns with $n_V \ge 2$ are related to the first column by a shift in the index $k$. The column $n_V=0$ can be obtained from normalization. Proceding by induction, the elements in the triangle can be finally recast in the compact formula
\begin{equation}
P_n(n_V|0)= \binom{n-n_V-1}{\lceil\frac{n-n_V-1}{2}\rceil} \binom{n_V}{\lceil\frac{n_V}{2}\rceil}2^{-n}.
\label{discrete_arcsine}
\end{equation}
Note that our result is slightly different from~\cite{feller}, where collisions are counted in pairs. When both $n$ and $n_V$ are large, we obtain the limit curve $P_n(n_V|0) \simeq 1/\sqrt{n_V(n-n_V)} \pi$.

When the scattering probability can vary in $0 \le p_s \le 1$, the triangle can be generated as above, and the first few terms read\\

    \begin{tabular}{ c| c c c c c c}
     $n$ & $n_V=0$ & $1$ & $2$ & $3$ & $4$\\
     \hline
     $0$ & $1$ & & & &\\
     $1$ & $\frac{1}{2}$ & $\frac{1}{2}$ & & &\\
     $2$ & $\frac{2-p_s}{4}$ & $\frac{2-p_s}{4}$ & $\frac{2p_s}{4}$ & &\\
     $3$ & $\frac{4-2p_s}{8}$ & $\frac{4-2p_s-p_s^2}{8}$ & $\frac{2p_s(2-p_s)}{8}$ & $\frac{3p_s^2}{8}$ &\\
     $4$ & $\frac{8-4p_s-p_s^3}{16}$ & $\frac{8-4p_s-2p_s^2}{16}$ & $\frac{2p_s(4-2p_s-p_s^2)}{16}$ & $\frac{3p_s^2(2-p_s)}{16}$ &  $\frac{6p_s^3}{16}$\\
    \end{tabular}
\vspace{.5cm}

Now, the identification of the polynomial coefficients $P_n(n_V|0)$ becomes more involved, because each coefficient is itself a polynomial with respect to $p_s$. The strategy in the identification is the same as above. By induction, the column for $n_V=1$ can be identified as being
\begin{eqnarray}
P_n(1|0)=\frac{1-p_s+\sqrt{1-p_s^2}}{4}+\nonumber \\
+\left( \frac{p_s}{2}\right)^{4+2y} \binom{2+2y}{1+y}\frac{~_2F_1\left(1, \frac{3}{2}+y, 3 + y, p_s^2 \right) }{2(2+y)} 
\end{eqnarray}
for $n \ge 2$, with $y=\lceil(n - 3)/2\rceil$, and $P_1(1|0)=1/2$. Once $P_n(1|0)$ is known, by inspection one realizes that the other columns $P_n(n_V|0)$ are related by
\begin{eqnarray}
P_n(n_V|0)= \left( \frac{p_s}{2}\right) ^{n_V-1} \binom{n_V}{\lceil\frac{n_V}{2}\rceil} P_{n-n_V+1}(1|0)
\end{eqnarray}
for $n_V \ge 2$. The probability $P_n(0|0)$ is finally obtained from normalization, and reads
\begin{eqnarray}
P_n(0|0)= \frac{p_s-1+\sqrt{1-p_s^2}}{2p_s}+\nonumber \\
+\left( \frac{p_s}{2}\right) ^{2+2z}\binom{2z}{z}\frac{~_2F_1\left(\frac{1}{2} + z, 1, 2 + z, p_s^2 \right)}{p_s(1+z)},
\end{eqnarray}
with $z=\lceil n /2\rceil$. These results generalize Eq.~\eqref{discrete_arcsine}, and are illustrated in Fig.~\ref{fig2}, where we compare $P_n(n_V|0)$ as a function of $n_V$ for $n=10$ and different values of $p_s$. When $p_s=1$ the distribution approaches a U shape, as expected. As soon as $p_s < 1$, the shape changes considerably, and in particular $P_n(n_V|0)$ becomes strongly peaked at $n_V=0$ as the effects of absorption overcome scattering. The presence of a second peak at $n_V=n$ is visible when $p_s \simeq 1$ and progressively disappear as $p_s$ decreases: when $p_s$ is small, $P_n(n_V|0)$ has an exponential tail. When $n$ is large, $P_n(n_V|0)$ approaches the asymptotic curve
\begin{equation}
P_\infty(n_V|0)=\left( \frac{p_s}{2}\right) ^{n_V-1} \binom{n_V}{\lceil\frac{n_V}{2}\rceil}\frac{1-p_s+\sqrt{1-p_s^2}}{4}
\label{asy_disc}
\end{equation}
for $n_V \ge 1$, and $P_\infty(0|0)=(p_s-1+\sqrt{1-p_s^2})/2p_s$. Remark that when $p_s=1$ this means that the U shape for large $n$ collapses on the two extremes at $n_V=0$ and $n_V=n$. Equation~\eqref{asy_disc} is an excellent approximation of $P_n(n_V|0)$ when the scattering probability is not too close to $p_s \simeq 1$: as expected, the discrepancy between the exact and aymptotic probability is most evident when $n_V \simeq n$, as shown in Fig.~\ref{fig2}. Fig.~\ref{fig3} displays $P_n(n_V|0)$ as a function of $n_V$ for $n=50$, in order to emphasize the effects of $p_s$: when $p_s=1$ the probability $P_n(n_V|0)$ is almost superposed to the asymptotic curve $P_n(n_V|0) \simeq 1/\sqrt{n_V(n-n_V)} \pi$, whereas a deviation in the scattering probability as small as $p_s = 0.95$ is sufficient to radically change the shape of the collision number distribution. Finally, observe that when $n_V$ is also large, which implies $p_a \ll 1$, Eq.~\eqref{asy_disc} behaves as
\begin{equation}
P_\infty(n_V|0) \simeq \sqrt{\frac{1-p_s}{\pi n_V}} e^{-(1-p_s) n_V}.
\label{exponential_asy}
\end{equation}

\begin{figure}[t]
   \centerline{\epsfclipon \epsfxsize=9.0cm
\epsfbox{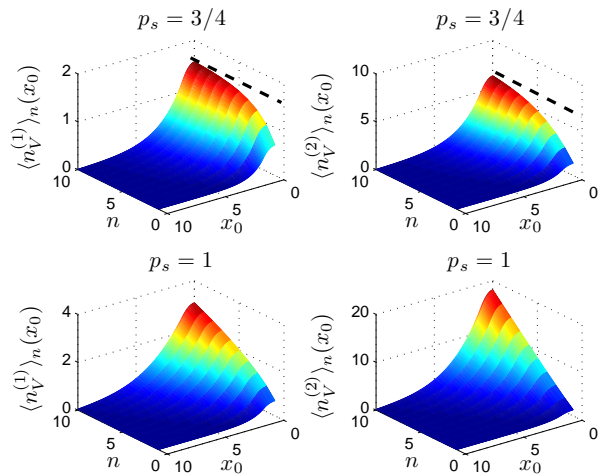} }
   \caption{(Color online). One-dimensional exponential flights in $V=[-1,1]$. Upper half. First (left) and second factorial moment (right) of $n_V$ as a function of $n$ and $x_0$, with $p_s=3/4$. The asymptotic values for $x_0=0$ are shown as a dahsed lines for reference. Lower half. First (left) and second factorial moment (right) of $n_V$ in the same domain, when $p_s=1$.}
   \label{fig6}
\end{figure}

\subsection{The arcsine law with continuous jumps}

When the displacement kernel $T(x' \to x)$ is continuous and symmetric, $p_s=1$, and $x_0=0$, the distribution of the number of collisions $n_V$ falling in $x\ge x_0$ is universal, in that it does not depend on the specific functional form of $T(x' \to x)$ (see~\cite{majumdar_fptd} and references therein). This strong and surprising result stems from a Sparre Andersen theorem~\cite{sparre}, whose proof is highly non trivial (and does not apply to discrete jumps)~\cite{majumdar_fptd}. This leaves the choice on the form of kernel $T(x' \to x)$, as far as it satisfies the hypotheses of the theorem. For the sake of simplicity, we have assumed an exponential distribution of jump lengths, i.e., $T(x' \to x)=s\exp(-s|x-x'|)/2$, with $s=1$. Starting from Eq.~\eqref{equation_F} and Eq.~\eqref{feynman_x0} we can generate the infinite triangle\\

    \begin{tabular}{ c| c c c c c c c c}
     $n$ & $n_V=0$ & $1$ & $2$ & $3$ & $4$ & $5$ & $6$ & $7$\\
     \hline
     $0$ & $1$ & & & & & & &\\
     $1$ & $\frac{1}{2}$ & $\frac{1}{2}$ & & & & & &\\
     $2$ & $\frac{3}{8}$ & $\frac{2}{8}$ & $\frac{3}{8}$ & & & & &\\
     $3$ & $\frac{3}{16}$ & $\frac{5}{16}$ & $\frac{5}{16}$ & $\frac{3}{16}$ & & & &\\
     $4$ & $\frac{35}{128}$ & $\frac{20}{128}$ & $\frac{18}{128}$ & $\frac{20}{128}$ & $\frac{35}{128}$ & & &\\
     $5$ & $\frac{63}{256}$ & $\frac{35}{256}$ & $\frac{30}{256}$ & $\frac{30}{256}$ & $\frac{35}{256}$ & $\frac{63}{256}$ & &\\
     $6$ & $\frac{231}{1024}$ & $\frac{126}{1024}$ & $\frac{105}{1024}$ & $\frac{100}{1024}$ & $\frac{105}{1024}$ & $\frac{126}{1024}$ & $\frac{231}{1024}$ &\\
     $7$ & $\frac{429}{2048}$ & $\frac{231}{2048}$ & $\frac{189}{2048}$ & $\frac{175}{2048}$ & $\frac{175}{2048}$ & $\frac{189}{2048}$ & $\frac{231}{2048}$ & $\frac{429}{2048}$\\
    \end{tabular}
\vspace{.5cm}

It is easy to verify that the triangle indeed does not depend on $s$, and that other functional forms of $T(x' \to x)$ would lead to the same coefficients for the polynomials $P_n(n_V|0)$. This holds true also for L\'evy flights, where $T(x' \to x)$ is a L\'evy stable law and jump lengths are unbounded~\cite{majumdar_fptd}. We start from the column $n_V=1$, observe the relation with the subsequent columns $n_V \ge 2$, and finally derive the case $n_V=0$ from normalization. Proceding therefore by induction we recognize that the elements of the triangle obey
\begin{equation}
P_n(n_V|0)= \binom{2n-2n_V}{n-n_V} \binom{2n_V}{n_V}2^{-2n}.
\label{satya_exact}
\end{equation}
We recover here the celebrated results of the collision number distribution for discrete-time walks with symmetric continuous jumps, in absence of absorption~\cite{feller, majumdar_fptd}. When both $n$ and $n_V$ are large, it is possible to show that Eq.~\eqref{satya_exact} converges to the U shape $1/\sqrt{n_V(n-n_V)} \pi$.

When the scattering probability is allowed to vary in $0 \le p_s \le 1$, it turns out that the polynomial coefficients $P_n(n_V|0)$ are the same for several different continuous symmetric kernels $T(x' \to x)$ (L\'evy flights included), and we are therefore led to conjecture that the universality result for the case $p_s=1$ carries over to to random walks with absorption. This allows generalizing the Sparre Andersen theorem for the collision number distribution on the half-line to a broader class of Markovian discrete-time processes. The first few terms in the triangle (which for practical purposes we have generated by resorting to $T(x' \to x)=s\exp(-s|x-x'|)/2$, with $s=1$) read\\

\begin{widetext}
    \begin{tabular}{ c| c c c c c c c}
     $n$ & $n_V=0$ & $1$ & $2$ & $3$ & $4$ & $5$\\
     \hline
     $0$ & $1$ & & & & &\\
     $1$ & $\frac{1}{2}$ & $\frac{1}{2}$ & & & &\\
     $2$ & $\frac{4-p_s}{8}$ & $\frac{4-2p_s}{8}$ & $\frac{3p_s}{8}$ & & &\\
     $3$ & $\frac{8-2p_s-p_s^2}{16}$ & $\frac{8-4p_s-p_s^2}{16}$ & $\frac{3p_s(2-p_s)}{16}$ & $\frac{5p_s^2}{16}$ & &\\
     $4$ & $\frac{64-16p_s-8p_S^2-5p_s^3}{128}$ & $\frac{64-32p_s-8p_s^2-4p_s^3}{128}$ & $\frac{6p_s(8-4p_s-p_s^2)}{128}$ & $\frac{20 p_s^2(2-p_s)}{128}$ &  $\frac{35 p_s^3}{128}$ &\\
$5$ & $\frac{128-32p_s-16p_s^2-10p_s^3-7p_s^4}{256}$ & $\frac{128-64p_s-16p_s^2-8p_s^3-5p_s^4}{256}$ & $\frac{6p_s(16-8p_s-2p_s^2-p_s^3)}{256}$ & $\frac{10 p_s^2(8-4p_s-p_s^2)}{256}$ & $\frac{35 p_s^3(2-p_s)}{256}$ & $\frac{63 p_s^4}{256}$\\
    \end{tabular}
\vspace{.5cm}
\end{widetext}

As above, identification of the terms $P_n(n_V|0)$ becomes more involved, because each coefficient is itself a polynomial with respect to $p_s$. By induction, the column for $n_V=1$ can be identified as being
\begin{eqnarray}
P_n(1|0)=\frac{\sqrt{1-p_s}}{2}+\nonumber \\
+\left(\frac{p_s}{4} \right)^n \binom{2n-2}{n-1} \frac{~_2F_1\left(-\frac{1}{2} + n, 1, 
   1 + n, p_s\right)}{n} 
\end{eqnarray}
for $n \ge 1$. Once $P_n(1|0)$ is known, the subsequent columns $P_n(n_V|0)$ are observed to obey
\begin{eqnarray}
P_n(n_V|0)= \left( \frac{p_s}{4}\right) ^{n_V-1} \binom{2n_V-1}{n_V} P_{n-n_V+1}(1|0)
\end{eqnarray}
for $n_V \ge 2$. The probability $P_n(0|0)$ is finally obtained from normalization, and reads
\begin{eqnarray}
P_n(0|0)= \frac{p_s-1+\sqrt{1-p_s}}{p_s}+\nonumber \\
+\left(\frac{p_s}{4} \right)^n \binom{2n}{n} \frac{~_2F_1\left(\frac{1}{2} + n, 1, 
   2 + n, p_s\right)}{2(1+n)}.
\end{eqnarray}
These results are illustrated in Fig.~\ref{fig4}, where we compare $P_n(n_V|0)$ as a function of $n_V$ for $n=10$ and different values of $p_s$. The findings for continuous jumps closely resemble those for discrete displacements. When $p_s=1$ the distribution approaches a U shape, as expected. As soon as $p_s < 1$, the shape again changes abruptly, and in particular $P_n(n_V|0)$ becomes strongly peaked at $n_V=0$ when absorption dominates scattering. When $p_s$ is small, $P_n(n_V|0)$ decreases exponentially at large $n_V$. When $n$ is large, $P_n(n_V|0)$ approaches the asymptotic curve
\begin{equation}
P_\infty(n_V|0)=\left( \frac{p_s}{4}\right) ^{n_V-1} \binom{2n_V-1}{n_V}\frac{\sqrt{1-p_s}}{2}
\label{asy_cont}
\end{equation}
for $n_V \ge 1$, and $P_\infty(0|0)=(p_s-1+\sqrt{1-p_s})/p_s$. Again, Eq.~\eqref{asy_cont} is an excellent approximation of $P_n(n_V|0)$ when the scattering probability is not too close to $p_s \simeq 1$, as shown in Fig.~\ref{fig4}. Fig.~\ref{fig5} displays $P_n(n_V|0)$ as a function of $n_V$ for $n=50$, in order to emphasize the effects of $p_s$: when $p_s=1$ the probability $P_n(n_V|0)$ is almost superposed to the asymptotic curve $P_n(n_V|0) \simeq 1/\sqrt{n_V(n-n_V)} \pi$, whereas a deviation in the scattering probability as small as $p_s = 0.95$ is sufficient to radically change the shape of the collision number distribution. Finally, observe that when $n_V$ is also large, which implies $p_a \ll 1$, Eq.~\eqref{asy_cont} yields the same scaling as Eq.~\eqref{exponential_asy}. All analytical calculations discussed here have been verified by comparison with Monte Carlo simulations with $10^6$ particles.

\section{Moments formulae}
\label{moments_formulae}

A complementary tool for characterizing the distribution $P_n(n_V|{\mathbf r}_0)$ is provided by the analysis of its moments. To this aim, it is convenient to introduce the function $\tilde{G}_n(u|{\mathbf r}_1)=\tilde{F}_n(1/u|{\mathbf r}_1)$. By construction, $\tilde{G}_n(u|{\mathbf r}_1)$ is the (rising) factorial moment generating function for trajectories entering their first collision at ${\mathbf r}_1$, which implies
\begin{equation}
\langle \tilde{n}_V^{(m)} \rangle_n({\mathbf r}_1) = (-1)^m \frac{\partial^m}{\partial u^m} \tilde{G}_n(u|{\mathbf r}_1) \vert_{u=1},
\label{recursion_mom_disc}
\end{equation}
$x^{(k)}=x(x+1)...(x+k-1)$ being the rising factorial~\cite{lah}. The tilde is used to recall that the moments refer to trajectories starting from the first collision at ${\mathbf r}_1$. Combining Eqs.~\eqref{equation_F} and~\eqref{recursion_mom_disc} yields the recursion property
\begin{eqnarray}
\langle \tilde{n}_V^{(m)} \rangle_{n+1}({\mathbf r}_1)-p_s\int  T^*({\mathbf r}'\to {\mathbf r}_1) \langle \tilde{n}_V^{(m)} \rangle_n({\mathbf r}')d{\mathbf r}'=\nonumber \\
=mV({\mathbf r}_1)\langle \tilde{n}_V^{(m-1)} \rangle_{n+1}({\mathbf r}_1),
\label{recursion_F_n}
\end{eqnarray}
for $m \ge 1$, with the conditions $\langle \tilde{n}_V^{(0)} \rangle_n({\mathbf r}_1)=1$, and $\langle \tilde{n}_V^{(m)} \rangle_1({\mathbf r}_1)=m!V({\mathbf r}_1)$. Finally, the factorial moments $\langle n_V^{(m)} \rangle_n({\mathbf r}_0)$ for particles emitted at ${\mathbf r}_0$ are obtained from
\begin{equation}
\langle n_V^{(m)} \rangle_n({\mathbf r}_0) = \int \langle \tilde{n}_V^{(m)} \rangle_n({\mathbf r}_1) T({\mathbf r}_0 \to {\mathbf r}_1)d{\mathbf r}_1.
\label{moments_T_r0}
\end{equation}

When trajectories are observed up to $n\to +\infty$, we can set $\langle n_V^{(m)} \rangle=\lim_{n\to +\infty}\langle n_V^{(m)} \rangle_n$, and from Eq.~\eqref{recursion_F_n} we find
\begin{eqnarray}
\langle \tilde{n}_V^{(m)} \rangle({\mathbf r}_1)-p_s\int  T^*({\mathbf r}'\to {\mathbf r}_1) \langle \tilde{n}_V^{(m)} \rangle({\mathbf r}')d{\mathbf r}'=\nonumber \\
=m  V({\mathbf r}_1) \langle \tilde{n}_V^{(m-1)} \rangle({\mathbf r}_1),
\label{L_moments_disc}
\end{eqnarray}
for $m \ge 1$, provided that $\langle \tilde{n}_V^{(m)} \rangle({\mathbf r}_1)$ is finite. It turns out that the asymptotic moments $\langle \tilde{n}_V^{(m)} \rangle({\mathbf r}_1)$ are related to the equilibrium distribution of the particles~\cite{zdm_prl, pitman, iosifescu}. To see this, we introduce the incident propagator $\Psi_n({\mathbf r}|{\mathbf r}_0)$, i.e., the probability density of finding a particle emitted at ${\mathbf r}_0$ entering the $n$-th collision ($n \ge 1$) at ${\mathbf r}$. We have
\begin{eqnarray}
\Psi_{n+1}({\mathbf r}|{\mathbf r}_0)=p_s\int  T({\mathbf r}'\to {\mathbf r}) \Psi_n({\mathbf r}'|{\mathbf r}_0)d{\mathbf r}',
\label{eq_psi}
\end{eqnarray}
with $\Psi_1({\mathbf r}|{\mathbf r}_0)=T({\mathbf r}_0 \to {\mathbf r})$. We introduce then the incident collision density
\begin{eqnarray}
\Psi({\mathbf r}|{\mathbf r}_0)=\lim_{N \to \infty} \sum_{n=1}^N \Psi_n({\mathbf r}|{\mathbf r}_0),
\end{eqnarray}
which can be interpreted as the particle equilibrium distribution~\cite{zdm_prl, zdm_pre, spanier, lux}. Now, by making use of the formal Neumann series~\cite{spanier}, from Eq.~\eqref{eq_psi} it follows that the collision density satisfies the stationary integral transport equation
\begin{eqnarray}
\Psi({\mathbf r}|{\mathbf r}_0)=p_s\int T({\mathbf r}'\to {\mathbf r})\Psi({\mathbf r}'|{\mathbf r}_0)d{\mathbf r}'+
T({\mathbf r}_0\to {\mathbf r}).
\label{integral_transport_equations}
\end{eqnarray}
The transport equation~\eqref{integral_transport_equations} can be understood as follows: at equilibrium, the stationary particle density entering a collision at ${\mathbf r}$ for a source emitting at ${\mathbf r}_0$ is given by the sum of all contributions entering a collision at ${\mathbf r}'$, being scattered and then transported to ${\mathbf r}$, plus the contribution of the particles emitted from the source and never collided up to entering ${\mathbf r}$. Now, by resorting to the relation between $T$ and its adjoint $T^*$ in Eq.~\eqref{scalar_product}, and observing that at equilibrium (for isotropic source and scattering)
\begin{equation}
\int T({\mathbf r}_0\to {\mathbf r}') \Psi({\mathbf r}|{\mathbf r}') d{\mathbf r}'=\int \Psi({\mathbf r}'|{\mathbf r}_0)T({\mathbf r}'\to {\mathbf r})d{\mathbf r}',
\label{equilibrium_balance}
\end{equation}
then Eq.~\eqref{L_moments_disc} can be inverted (see Appendix~\ref{appendix_integral}), and gives
\begin{eqnarray}
\langle \tilde{n}_V^{(m)} \rangle({\mathbf r}_1) = m p_s \int_V \Psi({\mathbf r}'|{\mathbf r}_1) \langle \tilde{n}_V^{(m-1)} \rangle({\mathbf r}')d{\mathbf r}' +\nonumber \\
+mV({\mathbf r}_1)\langle \tilde{n}_V^{(m-1)} \rangle({\mathbf r}_1),
\label{factorial_recursion_psi}
\end{eqnarray}
for $m \ge 1$, where $\Psi$ is the solution of Eq.~\eqref{integral_transport_equations}. Thus, by using Eqs.~\eqref{moments_T_r0} and~\eqref{equilibrium_balance}, from Eq.~\eqref{factorial_recursion_psi} it follows
\begin{equation}
\langle n_V^{(m)} \rangle({\mathbf r}_0) = m \int_V \Psi({\mathbf r}'|{\mathbf r}_0) \langle \tilde{n}_V^{(m-1)} \rangle({\mathbf r}') d{\mathbf r}'.
\end{equation}
Hence, by induction we finally get the desired relation between the factorial moments and the equilibrium distribution, namely,
\begin{equation}
\langle n_V^{(m)} \rangle({\mathbf r}_0) = \sum_{k=1}^{m} L_{m,k} p_s^{k-1} C_k({\mathbf r}_0),
\label{asy_moments_formula}
\end{equation}
where
\begin{eqnarray}
C_k({\mathbf r}_0)=k!\int_V \cdots \int_V \prod_{i=1}^{k}d{\mathbf r}_i \Psi({\mathbf r}_i|{\mathbf r}_{i-1})
\end{eqnarray}
are $k$-fold convolution (Kac) integrals over the equilibrium distribution $\Psi({\mathbf r}|{\mathbf r}_{0}) $, and the coefficients $L_{m,k}=\binom {m} {k} (m-1)!/(k-1)!$ are the Lah numbers~\cite{lah}. Observe that a result analogous to Eq.~\eqref{asy_moments_formula} has been derived in~\cite{zdm_prl} building upon survival probabilities.

\section{First and second moment of collision number: examples of calculations}
\label{applications_moments}

To illustrate the approach proposed in the previous Section, we compute here the average collision number $\langle n_V^{(1)} \rangle_n(x_0)$ and the second factorial moment $\langle n_V^{(2)} \rangle_n(x_0)$ in a bounded domain $V$ in one dimension. We choose an exponential displacement kernel $T(x' \to x)=s\exp(-s|x-x'|)/2$, with $s=1$, which is often adopted as a simplified random walk model (the so-called exponential flights) to describe gas dynamics, radiation propagation or biological species migration~\cite{mazzolo, wing, bacteria, velocity_jump, weiss_review, lorentz}. In this respect, the average $\langle n_V^{(1)} \rangle_n(x_0)$ is a measure of the passage of the particles through the region $V$ (the deposited energy, for instance), whereas the second moment $\langle n_V^{(2)} \rangle_n(x_0)$ is proportional to the incertitude on the average~\cite{zdm_prl, zdm_pre}.

The calculations stemming from Eqs.~\eqref{recursion_F_n} and~\eqref{moments_T_r0} are rather cumbersome, so that it is preferable to visually represent our results instead of writing down the explicit formulas. In Fig.~\ref{fig6} we display the behavior of the moments $\langle n_V^{(1)} \rangle_n(x_0)$ and $\langle n_V^{(2)} \rangle_n(x_0)$ for the volume $V$ being the interval $[-R,R]$, with $R=1$. When $p_s=1$, the moments diverge as $n$ increases, since exponential flights in one dimension are recurrent random walks, and revisit their starting point infinitely many times. When $p_s <1$, they converge instead to an asymptotic value, which can be computed based on Eq.~\eqref{asy_moments_formula} by observing that the collision density for this example is
\begin{equation}
\Psi(x|x_0)=\frac{e^{-\sqrt{1-p_s}|x-x_0|}}{2\sqrt{1-p_s}},
\end{equation}
as discussed in~\cite{zdm_prl}. Moreover, Fig.~\ref{fig6} shows that the moments decrease as the distance of the source $x_0$ from the region $V$ increases, as expected. As we have chosen here a symmetric interval, we have $\langle n_V^{(m)} \rangle_n(x_0)=\langle n_V^{(m)} \rangle_n(-x_0)$, so that we can plot the moments only for positive values of $x_0$. All results presented in this Section have been verified by comparison with Monte Carlo simulations with $10^6$ particles. Other kinds of boundary conditions (leakage, for instance, which implies that particles are lost upon crossing the frontier of $V$~\cite{zdm_prl, zdm_pre}) have also been successfully tested, but will not be presented here.

\section{Diffusion limit}
\label{diffusion}

To conclude our analysis, in this Section we comment on the scaling limit of the discrete Feynman-Kac equation, which is achieved when $n_V$ is large, and at the same time the typical jump length $\epsilon$ is vanishing small. We set $t_V=n_V dt$ and $t=n dt$, where $dt$ is some small time scale, related to $\epsilon$ by the usual diffusion scaling $\epsilon^2 = 2D dt$, the constant $D$ playing the role of a diffusion coefficient. When $T$ is not symmetric, so that displacements have mean $\mu$, we further require $\mu = v dt $, where the constant $v$ is a velocity. By properly taking the limit of large $n_V$ and vanishing $dt$, $t_V$ converge to the residence time in $V$. The quantity $n_V$ can only be large if the absorption probability $p_a$ is small, and it is natural to set $p_a=\lambda_a dt$, the quantity $\lambda_a$ being an absorption rate per unit of $dt$. Observe that when both $\epsilon$ and $\mu$ are small for any displacement kernel we have the Taylor expansion
\begin{equation}
\int  T({\mathbf r}'\to {\mathbf r})f({\mathbf r}')d{\mathbf r}' \simeq f( {\mathbf r}) - \mu \partial_{{\mathbf r}}f( {\mathbf r}) +\frac{1}{2}\epsilon^2\partial^2_{{\mathbf r}}f( {\mathbf r}),
\end{equation}
where the first-order derivative vanishes if the kernel is symmetric. A similar expansion holds for the kernel $T^*$, namely,
\begin{equation}
\int  T^*({\mathbf r}'\to {\mathbf r}_0)f({\mathbf r}')d{\mathbf r}' \simeq f( {\mathbf r}_0)+ \mu \partial_{{\mathbf r}_0}f( {\mathbf r}_0) + \frac{1}{2}\epsilon^2\partial^2_{{\mathbf r}_0}f( {\mathbf r}_0).
\end{equation}
It is expedient to introduce the quantity $Q_t(u|{\mathbf r}_0)=F_t(e^{-u}|{\mathbf r}_0)$, which is the moment generating function of $t_V=n_V dt$, i.e.,
\begin{equation}
\langle t_V^{m} \rangle_t({\mathbf r}_0) = (-1)^m \frac{\partial^m}{\partial u^m} Q_t(u|{\mathbf r}_0) \vert_{u=0},
\label{recursion_diffusion}
\end{equation}
when trajectories are observed up to $t=n dt$. Under the previous hypotheses, combining Eqs.~\eqref{equation_F} and~\eqref{feynman_x0} yields
\begin{eqnarray}
Q_{t+dt}(u|{\mathbf r}_0)-Q_{t}(u|{\mathbf r}_0)\simeq\nonumber \\
\simeq{\cal L}_{{\mathbf r}_0}^*Q_{t}(u|{\mathbf r}_0)dt-uV({\mathbf r}_0)Q_{t}(u|{\mathbf r}_0)dt+\lambda_a dt,
\end{eqnarray}
where we have neglected all terms vanishing faster than $dt$, and ${\cal L}_{{\mathbf r}_0}^*=D\partial^2_{{\mathbf r}_0}+v\partial_{{\mathbf r}_0}-\lambda_a$. Taking the limit $dt \to 0$ we recognize then the Feynman-Kac equation for a Brownian motion with diffusion coefficient $D$, drift $v$ and absorption rate $\lambda_a$, namely
\begin{eqnarray}
\frac{\partial Q_{t}(u|{\mathbf r}_0)}{\partial t}={\cal L}_{{\mathbf r}_0}^*Q_{t}(u|{\mathbf r}_0)-uV({\mathbf r}_0)Q_{t}(u|{\mathbf r}_0)+\lambda_a.
\label{eq_feynman_diffusion_dt}
\end{eqnarray}
In other words, in the diffusion limit the statistical properties of the hit number in $V$ behave as those of the residence time of a Brownian motion, as is quite naturally expected on physical grounds~\cite{berezhkovskii, benichou_epl, zdm_pre}. Finally, from Eq.~\eqref{recursion_diffusion} stems the recursion property for the moments
\begin{equation}
\frac{\partial \langle t_V^{m} \rangle_t({\mathbf r}_0)}{\partial t} = {\cal L}_{{\mathbf r}_0}^*\langle t_V^{m} \rangle_t({\mathbf r}_0) + mV({\mathbf r}_0)\langle t_V^{m-1} \rangle_t({\mathbf r}_0) ,
\end{equation}
in agreement with the results in~\cite{agmon, agmon_lett} for Brownian motion.

\section{Conclusions}
\label{conclusions}

In this paper we have examined the behavior of the distribution $P_n(n_V|{\mathbf r}_{0})$ of the number of hits $n_V$ in a region $V$ for a broad class of stochastic processes with scattering and absorption. Key to our analysis has been a discrete version of the Feynman-Kac formalism. We have shown that this approach is amenable to explicit formulas for $P_n(n_V|{\mathbf r}_{0})$, at least for simple geometries and displacement kernels. The moments of the distribution have also been detailed, and their asymptotic behavior for large $n$ has been related to the walker equilibrium density. Finally, the diffusion limit and the convergence to the Feynman-Kac formulae for Brownian motion have been discussed.

We conclude by observing that a generalization of the present work to more realistic transport kernels, including anisotropic source and scattering, would be possible, for instance by resorting to the formalism proposed in~\cite{zdm_pre_operator}. Moreover, while in this paper we have focused on counting statistics, and therefore chosen $V({\mathbf r})$ to be the marker function of a given domain in phase space, the Feynman-Kac formalism can be adapted with minor changes to describing the statistics of other kinds of functionals, such as for instance hitting probabilities~\cite{barkai_jsp, redner, zoia_rosso_majumdar}.

\appendix

\section{The stationary moment equation}
\label{appendix_integral}

We want to solve an integral equation
\begin{equation}
f({\mathbf r}_1) - p_s \int T^*({\mathbf r}'\to {\mathbf r}_1)f({\mathbf r}')d{\mathbf r}'=g({\mathbf r}_1)
\label{integral_eq}
\end{equation}
for the function $f({\mathbf r}_1)$, where $g({\mathbf r}_1)$ is known. We propose a solution in the form
\begin{equation}
f({\mathbf r}_1) = p_s \int g({\mathbf r}') \Psi({\mathbf r}'|{\mathbf r}_1)d{\mathbf r}'+g({\mathbf r}_1)
\label{integral_sol}
\end{equation}
and ask which is the equation satisfied by the integral kernel $\Psi({\mathbf r}'|{\mathbf r}_1)$. By injecting Eq.~\eqref{integral_sol} into Eq.~\eqref{integral_eq} one obtains
\begin{eqnarray}
\int g({\mathbf r}') \Psi({\mathbf r}'|{\mathbf r}_1) d{\mathbf r}' =\nonumber \\
= p_s \int \int \Psi({\mathbf r}''|{\mathbf r}') T^*({\mathbf r}'\to {\mathbf r}_1)  g({\mathbf r}'')d{\mathbf r}' d{\mathbf r}''+\nonumber \\
+\int T^*({\mathbf r}'\to {\mathbf r}_1)  g({\mathbf r}')d{\mathbf r}'.
\label{integral_combi}
\end{eqnarray}
Recall that the adjoint and direct displacement kernels are related to each other by the scalar products
\begin{eqnarray}
\int g({\mathbf r})\int T({\mathbf r}'\to {\mathbf r}) f({\mathbf r}') d{\mathbf r}' d{\mathbf r}= \nonumber \\
=\int f({\mathbf r})\int  T^*({\mathbf r}'\to {\mathbf r})g({\mathbf r}') d{\mathbf r}' d{\mathbf r}
\label{scalar_product}
\end{eqnarray}
for any test functions $f$ and $g$~\cite{spanier}. From the definition of the scalar product in Eq.~\eqref{scalar_product} it follows that the second term at the right hand side of Eq.~\eqref{integral_combi} is given by
\begin{eqnarray}
\int T^*({\mathbf r}'\to {\mathbf r}_1)  g({\mathbf r}')d{\mathbf r}'=\int T({\mathbf r}_1\to {\mathbf r}')  g({\mathbf r}')d{\mathbf r}'.
\end{eqnarray}
From Eqs.~\eqref{scalar_product} and~\eqref{equilibrium_balance} the first term at the right hand side of Eq.~\eqref{integral_combi} becomes
\begin{eqnarray}
\int \int \Psi({\mathbf r}''|{\mathbf r}') T^*({\mathbf r}'\to {\mathbf r}_1)  g({\mathbf r}'')d{\mathbf r}' d{\mathbf r}'' = \nonumber \\
\int \int \Psi({\mathbf r}'|{\mathbf r}_1) T({\mathbf r}'\to {\mathbf r}'') g({\mathbf r}'')d{\mathbf r}' d{\mathbf r}''.
\end{eqnarray}
Therefore, $\Psi({\mathbf r}'|{\mathbf r}_1)$ obeys
\begin{eqnarray}
\int g({\mathbf r}') \Psi({\mathbf r}'|{\mathbf r}_1) d{\mathbf r}' =\nonumber \\
=p_s \int \int \Psi({\mathbf r}'|{\mathbf r}_1) T({\mathbf r}'\to {\mathbf r}'') g({\mathbf r}'')d{\mathbf r}' d{\mathbf r}''+\nonumber \\
+ \int T({\mathbf r}_1\to {\mathbf r}')  g({\mathbf r}')d{\mathbf r}',
\end{eqnarray}
which for the arbitrariness of $g({\mathbf r}')$ finally implies Eq.~\eqref{integral_transport_equations}, i.e., the required kernel $\Psi$ satisfies the integral transport equation.

\acknowledgments

The authors wish to thank Dr.~F.~Malvagi for useful discussions.


\begin{thebibliography}{10}
\bibitem{hughes} B.~D.~Hughes, {\em Random walks and random environments} Vol. I (Clarendon Press, Oxford, 1995).
\bibitem{weiss} G.~H.~ Weiss, {\em Aspects and applications of the random walk} (North Holland Press, Amsterdam, 1994).
\bibitem{bouchaud_desorder} J.~Ph.~Bouchaud and A.~Georges, Phys.~Rep.~{\bf 195}, 127 (1990).
\bibitem{avraham} D.~ben Avraham and S.~Havlin, {\em Diffusion and reactions in fractals and disordered systems} (CUP, UK, 2000).
\bibitem{redner} S.~Redner, {\em A guide to first-passage processes} (CUP, UK, 2001).
\bibitem{condamin_benichou} S.~Condamin et al., Nature {\bf 450}, 40 (2007).
\bibitem{klafter} R.~Metzler and J.~Klafter, Phys.~Rep.~{\bf 339}, 1 (2000).
\bibitem{condamin} S.~Condamin, O.~B\'enichou, and M.~Moreau, Phys.~Rev.~Lett.~{\bf 95}, 260601 (2005).
\bibitem{benichou_grebenkov} O.~B\'enichou et al., J.~Stat.~Phys.~{\bf 142}, 657 (2011).
\bibitem{barkai} E.~Barkai, J.~Stat.~Phys.~{\bf 123}, 883 (2006).
\bibitem{grebenkov} D.~S.~Grebenkov, Phys.~Rev.~E {\bf 76}, 041139 (2007).
\bibitem{grebenkov_jsp} D.~S.~Grebenkov, J.~Stat.~Phys.~{\bf 141}, 532 (2010).
\bibitem{majumdar_occupation} S.~N.~Majumdar and A.~Comtet, Phys.~Rev.~Lett.~{\bf 89}, 060601 (2002).
\bibitem{cercignani} C.~Cercignani, {\em The Boltzmann equation and its applications} (Springer, 1988).
\bibitem{wigner} M.~Weinberg and E.~P.~Wigner, {\em The physical theory of neutron chain reactors} (UCP, Chicago, 1958).
\bibitem{jacoboni_book} C.~Jacoboni and P.~Lugli, {\em The Monte Carlo method for semiconductor device simulation} (Springer, 1989).
\bibitem{gamma_biology} F.~Bartumeus et al., J.~Theor.~Bio.~{\bf 252}, 43 (2008).
\bibitem{lecaer} G.~Le Ca\"{e}r, J.~Stat. Phys.~{\bf 140}, 728 (2010). % 1d and 2d propagator
\bibitem{zoia} A.~Zoia, A.~Rosso, and S.~N.~Majumdar, Phys.~Rev.~Lett.~{\bf 102}, 120602 (2009).
\bibitem{schlesinger} M.~F.~Shlesinger, Nature {\bf 443}, 281 (2006).
\bibitem{kac_original} M.~Kac, Trans.~Amer.~Math.~Soc.~{\bf 65}, 1 (1949). % original
\bibitem{kac_berkeley} M.~Kac, in Proc.~Second Berkeley Symp.~on Math.~Statist.~and Prob.~(UCP, 1951), pp.~189-215.
\bibitem{kac_darling} D.~A.~Darling and M.~Kac, Trans.~Amer.~Math.~Soc.~{\bf 84}, 444 (1957). % markov
\bibitem{kac} M.~Kac, {\em Probability and related topics in physical sciences} (Lectures in applied mathematics, Wiley, 1957).
\bibitem{majumdar_review} S.~N.~Majumdar, Curr.~Sci.~{\bf 89}, 2076 (2005).
\bibitem{agmon_original} N.~Agmon, J.~Chem.~Phys.~{\bf 81}, 3644 (1984).% residence times
\bibitem{berezhkovskii} A.~M.~Berezhkovskii, V.~Zaloj, and N.~Agmon, Phys.~Rev.~E {\bf 57}, 3937 (1998).% residence times
\bibitem{benichou_epl} O.~B\'enichou et al., Europhys.~Lett.~{\bf 70}, 42 (2005).
\bibitem{agmon_lett} N.~Agmon, Chem.~Phys.~Lett.~{\bf 497}, 184 (2010).% residence times
\bibitem{agmon} N.~Agmon, J.~Phys.~Chem.~A {\bf 115}, 5838 (2011).% residence times
\bibitem{zoia_dumonteil_mazzolo} A.~Zoia, E.~Dumonteil, and A.~Mazzolo, Phys.~Rev.~E~{\bf 83}, 041137 (2011).
\bibitem{zdm_prl} A.~Zoia, E.~Dumonteil, and A.~Mazzolo, Phys.~Rev.~Lett.~{\bf 106}, 220602 (2011).
\bibitem{zdm_pre} A.~Zoia, E.~Dumonteil, and A.~Mazzolo, Phys.~Rev.~E~{\bf 84}, 021139 (2011).
\bibitem{turgeman} L.~Turgeman, S.~Carmi, and E.~Barkai, Phys.~Rev.~Lett.~{\bf 103}, 190201 (2009).
\bibitem{barkai_jsp} S.~Carmi, L.~Turgeman, and E.~Barkai, J.~Stat.~Phys.~{\bf 141}, 1071 (2010).
\bibitem{klafter_fptd} A.~Bar-Haim and J.~Klafter, J.~Chem.~Phys.~{\bf 109}, 5187 (1998).
\bibitem{delmoral} P.~Del Moral, {\em Feynman-Kac formulae. Genealogical and interacting particle systems with applications}, (Springer, NY, 2004).
\bibitem{pitman} P.~J.~Fitzsimmons and J.~Pitman, Stoch.~Proc.~Appl.~{\bf 79}, 117 (1999). % review markov
\bibitem{iosifescu} M.~Iosifescu, {\em Finite Markov processes and their applications}, (Wiley series in probability and mathematical statistics, Wiley, 1980).
\bibitem{csaki} E.~Cs\`{a}ki, J.~Stat.~Plann.~Inference {\bf 34}, 63 (1993).
\bibitem{blanco} S.~Blanco and R.~Fournier, Phys.~Rev.~Lett.~{\bf 97}, 230604 (2006). % short paths
\bibitem{spanier} J.~Spanier and E.~M.~Gelbard, {\em Monte Carlo principles and neutron transport problems} (Addison-Wesley, Reading, 1969).
\bibitem{lux} I.~Lux and L.~Koblinger, {\em Monte Carlo particle transport methods: neutron and photon calculations} (CRC Press, Boca Raton, 1991).
\bibitem{meyn} S.~P.~Meyn and R.~L.~Tweedie, Adv.~Appl.~Prob.~{\bf 24}, 542 (1992).
\bibitem{feller} W.~Feller, {\em An introduction to probability theory and its applications}, 3rd edition (Wiley, New York, 1970).
\bibitem{lamperti} J.~Lamperti, Trans.~Am.~Math.~Soc.~{\bf 88}, 380 (1958).
\bibitem{majumdar_potential} S.~Sabhapandit, S.~N.~Majumdar, and A.~Comtet, Phys.~Rev.~E {\bf 73}, 051102 (2006).
\bibitem{godreche} C.~Godr\`{e}che and J.~M.~Luck, J.~Stat.~Phys.~{\bf 104}, 489 (2001).
\bibitem{baldassarri} A.~Baldassarri et al., Phys.~Rev.~E {\bf 59}, 20 (1999).
\bibitem{watanabe} S.~Watanabe, Proc.~Symp.~Pure Math.~{\bf 57}, 157 (1995).
\bibitem{majumdar_fptd} S.~N.~Majumdar, Physica A {\bf 389}, 4299 (2010).
\bibitem{sparre} E.~Sparre Andersen, Math.~Scand.~{\bf 2}, 195 (1954).
\bibitem{lah} J.~Riordan, {\em Introduction to combinatorial analysis} (PUP, 1980).
\bibitem{mazzolo} A.~Mazzolo, Europhys.~Lett.~{\bf 68}, 350 (2004).
\bibitem{wing} G.~Milton Wing, {\em An introduction to transport theory} (Wiley, NY, 1962).
\bibitem{bacteria} H.~T.~Hillen and G.~Othmer, Siam J.~Appl.~Math {\bf 61}, 751 (2000).
\bibitem{velocity_jump} H.~G.~Othmer, S.~R.~Dunbar, and W.~Alt, J.~Math.~Biol.~{\bf 26}, 263 (1988).
\bibitem{weiss_review} G.~H.~ Weiss, Physica A {\bf 311}, 381 (2002).
\bibitem{lorentz} P.~L.~Krapivsky, S.~Redner, and E.~Ben-Naim, {\em A kinetic view of Statistical Physics} (CUP, UK, 2010).
\bibitem{zdm_pre_operator} A.~Zoia, E.~Dumonteil, and A.~Mazzolo, Phys.~Rev.~E, accepted.
\bibitem{zoia_rosso_majumdar} S.~N.~Majumdar, A.~Rosso, and A.~Zoia, Phys.~Rev.~Lett.~{\bf 104}, 020602 (2010).
\end{thebibliography}
\end{document}